# One dimensional photonic media characterized by uniform distributions of clusters

Michele Bellingeri[1], Francesco Scotognella[2]

[1] Dipartimento di Fisica, Università di Parma, Viale Usberti 7/A, I-43100, Parma, Italy

[2] Dipartimento di Fisica, Politecnico di Milano, P.zza Leonardo da Vinci 32, I-20133, Milano, Italy

**Abstract**

We realized one dimensional disordered photonic structures by grouping high refractive index layers in clusters, randomly distributed in such structure. We have control on the maximum size of the cluster and on the ratio high-low refractive index layers. By studying the average transmission as a function of the maximum cluster size, we observe a trend with a minimum and very interesting oscillations. Our results provide new insight in the physics of random systems.

**Introduction**

In the last years a large effort has been spent to study the optical properties of disordered photonic structures [1-4]. Such study have been catalysed by intriguing applications of these materials, such as random lasers [5-7]. In a photonic crystal, where the alternation of high and low refractive index materials in one, two and three dimensions is periodic [8-11], a laser can be obtained by exploiting the edges of a particular spectral region (namely, the photonic band gap), in which light is not allowed to pass through [12-15]. Instead, in a random photonic structure, with high and low refractive index materials randomly distributed [16-18], laser emission is observed in correspondence of random transmission depths [19-20].

In one dimension, disorder photonic systems are realized with a simple random alternation of high and low refractive index layers [21]. Although, the optical properties of these materials are very intriguing and have been intensively studied. In 2005, Bertolotti et al. experimentally observed optical necklaces [22], while in 2007 Ghulinyan observed periodic oscillations in the average transmission as a function of the sample length [23]. Recently, the Shannon index, a diversity index used in information theory and statistics [24], has been used to quantify the evenness in one dimensional photonic crystals [25,26].

In this paper, we realized one dimensional disordered photonic structures by grouping high refractive index layers in clusters, randomly distributed in such structure. We have control on the maximum size of the cluster and on the ratio between high and low refractive index layers (that we call dilution). We observe a trend with a minimum and very interesting oscillations by studying the average transmission as a function of the maximum cluster size.

**Methods**

We have realized for this study different photonic crystals with different sizes of the unit cell. As shown in Figure 1, the unit cell is made by one high refractive index layer. For each high refractive index layer, the unit cell contains $m$-1 low refractive index layer. $m$ is an index of the high refractive index layer dilution in the medium. For example, if $m$=10, 1 layer each 10 layers in the medium will be a high refractive index layer, and we refer to this structure as a 1/10 diluted structure. In this work, we have used different dilutions: 1/2, 1/4 , 1/6, 1/8, 1/10 and 1/12. For the different crystal structures we always have 100 high refractive index layers, such that the total number of layers for the structure is, respectively, 200 for $m$=2, 400 for $m$=4, 600 for $m$=6, 800 for $m$=8, 1000 for $m$=10, 1200 for $m$=12.

Then, corresponding to each photonic crystal, we have realized disordered structures by grouping the high refractive index layers in clusters. We sorted the clusters size from the uniform distribution in the interval (1, $k_{max}$), where $k_{max}$ is the maximum cluster size. Thus, increasing the parameter $k_{max}$, we increase the probability to have large high refractive layer clusters in the structure. Once the cluster distribution is computed, we randomly distributed these clusters in the structures. The $k_{max}$ parameter can be viewed as a way to control the homogeneity of the medium: the more $k_{max}$ is, the less the homogeneity of the medium is.

Thus all the realized crystal show the same number of refractive elements, but differ in the manner of aggregating the high refractive layers in clusters. The photonic crystal presents the high refractive layers with periodic distance among them, where the disordered structure show clusters of size related to the $k_{max}$ parameter and the distance among clusters is stochastic.

To give an example, a structure with $m$=4 and $k_{max}$=2 has 100 high refractive index layers, 300 low refractive index layers, and the clusters made with high refractive index layers have a maximum size of 2 layers. Differently, a structure with $m$=4 and $k_{max}$=5 has 100 high refractive and 300 low refractive index layers, but the clusters made with high refractive index layers have a maximum size of 5 layers. In this manner, the $k_{max}$=2 crystal is more uniform (more homogeneous) than the $k_{max}$=5 medium.

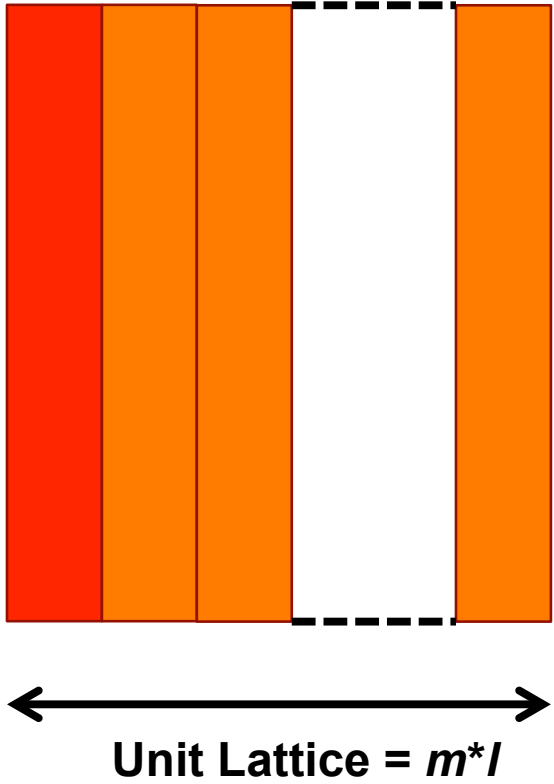

**Unit Lattice = *m*l***

**Figure 1.** Scheme of the photonic structure unit lattice (referring to corresponding periodic photonic crystal). For each high refractive index layer, the unit cell contains *m*-1 low refractive index layer. *m* is an index of the high refractive index layer dilution in the medium. For example, if *m*=10, 1 layer each 10 layers in the medium will be a high refractive index layer, and we refer to this structure as a 1/10 diluted structure.

Calculations of the light transmission through the media (in the spectral region corresponding to the photonic band gap of the periodic photonic crystal) are performed with the transfer matrix method [27]. We have considered isotropic, non-magnetic materials shaping the system glass/multilayer/air (in which glass is the sample substrate) and an incidence of the light normal to the stacked layer surface. $n_0$ and $n_S$ are the refractive indexes of air and glass, respectively, while $E_m$ and $H_m$ are the electric and magnetic fields in the glass substrate. To determine the electric and magnetic fields in air, $E_0$ and $H_0$, we have solved the following system:

$$\begin{bmatrix} E_0 \\ H_0 \end{bmatrix} = M_1 \cdot M_2 \cdot \ldots \cdot M_m \begin{bmatrix} E_m \\ H_m \end{bmatrix} = \begin{bmatrix} m_{11} & m_{12} \\ m_{21} & m_{22} \end{bmatrix} \begin{bmatrix} E_m \\ H_m \end{bmatrix} \qquad (1)$$

where

$$M_j = \begin{bmatrix} A_j & B_j \\ C_j & D_j \end{bmatrix},$$

with $j=(1,2,\ldots,m)$, is the characteristic matrix of each layers. The elements of the transmission matrix *ABCD* are

$$A_j = D_j = \cos(\phi_j), \ B_j = -\frac{i}{p_j}\sin(\phi_j), \ C_j = -i\, p_j \sin(\phi_j), \qquad (2)$$

where $n_j$ and $d_j$, hided in the angle $\phi_j$, are respectively the effective refractive index and the thickness of the layer *j*. In the case of normal incidence of the probe beam, the phase variation of the wave passing the *j*-fold layer is $\phi_j = (2\pi/\lambda)n_j d_j$, while the coefficient $p_j = \sqrt{\varepsilon_j/\mu_j}$ in TE wave and $q_j=1/p_j$ replace $p_j$ in TM wave. Inserting Equation (2) into Equation (1) and using the definition of transmission coefficient,

$$t = \frac{2p_s}{(m_{11} + m_{12}p_0)p_s + (m_{21} + m_{22}p_0)} \qquad (3)$$

it is possible to write the light transmission as

$$T = \frac{p_0}{p_s}|t|^2 \qquad (4)$$

**Results and Discussion**

In Figure 2 we show the transmission spectra of a periodic photonic crystal and of a disordered photonic crystal. In particular, the photonic crystal (blue curve in Figure 2) is made by 100 unit cells, and each cell contains a high refractive index layer (i.e. a total thickness of 70 nm) and seven low refractive index layers (i.e. a total thickness of 560 nm). In this work, we refer to such compounds as 1/8 diluted structures. Instead, the disordered photonic structure has the same number of layers of the periodic crystal, with a maximum cluster size $k_{max}$ of 4. While the periodic photonic crystal shows a photonic band gap, the disordered structure show narrow peaks overall the studied spectral region.

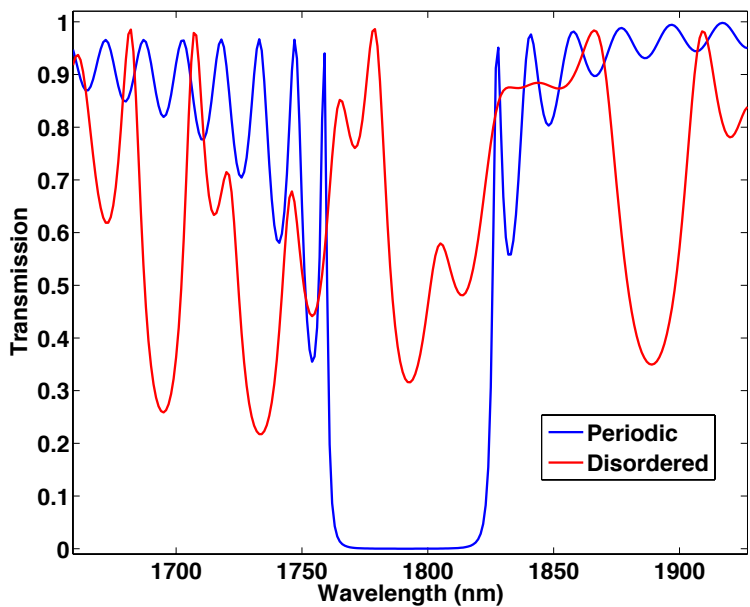

**Figure 2.** a) Transmission spectra of a periodic photonic crystal; the photonic crystal is made by 100 unit cell and each unit cell consists of 1 high refractive index layer and 7 low refractive index layers (we call it a crystal with 1/8 dilution). b) Transmission spectra of a disordered photonic structure with 1/8 dilution and with maximum cluster size $k_{max}$=4.

We calculated the total transmission of the disordered structures in the spectral region where the photonic band gap of the corresponding periodic structure is present. More precisely, we consider the full width at half maximum of the photonic band gap (in the case in Figure 2, 1/8 dilution, we consider the region 1759-1827 nm).

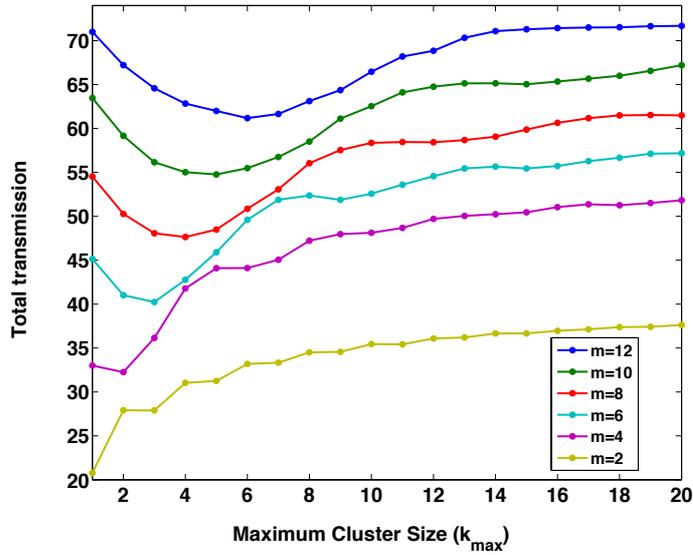

**Figure 3.** Total transmission as a function of the maximum cluster size $k_{max}$ (from 1 to 20) for the different cluster dilutions.

In Figure 3 it is evident the shift towards larger cluster size for the minimum of the curve. This can be due by the fact that the same level of non-homogeneity can be obtained, for larger dilutions, with larger $k_{max}$. This is ascribed to the fact that for larger dilution (lower high/low refractive index layer ratio) the light is more efficiently entrapped with the presence of larger clusters.

We observe interesting oscillations of the total transmission as a function of the maximum cluster size $k_{max}$. This phenomenon is more clear in Figure 4, where the total transmission for the 1/8 diluted disordered structure as a function of the maximum cluster size (up to $k_{max}$=30) is shown. We noticed, after the minimum corresponding to $k_{max}$=4, an increase of the total transmission and three oscillations with a similar period. This can be due to the concurrence of two effects: i) bigger clusters correspond to larger low refractive index regions and, consequently, large total transmittance for light; ii) we have cluster sizes that act as "higher orders" of the first total transmission minimum.

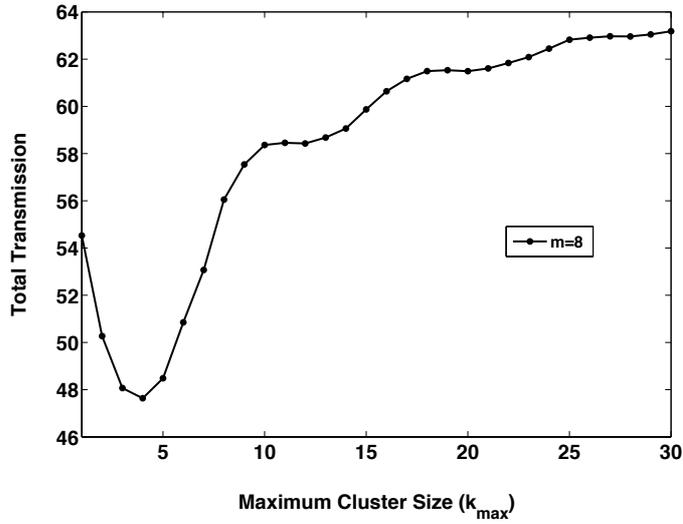

**Figure 4.** Total transmission as a function of the maximum cluster size $k_{max}$, with $1 \leq k_{max} \leq 30$, for 1/8 dilution.

## Conclusions

In this work we have realized one dimensional disordered photonic structures by grouping high refractive index layers in clusters, randomly distributed in such structure. We have control on the maximum size of the cluster and on the ratio between high and low refractive index layers. By studying the average transmission as a function of the maximum cluster size, we observe a trend with a minimum and very interesting oscillations. Our results provide new insight in the physics of random systems.

## Acknowledgement

We would like to thank Stefano Poletti for useful discussions.

.